\documentclass{iau}
\usepackage{graphicx}
\usepackage{color}
\usepackage{soul}
\title[SO/PHI] 
{The Polarimetric and Helioseismic Imager for {\em Solar Orbiter}: SO/PHI}

\author[S.K. Solanki and the SO/PHI team]   
{Sami K. Solanki$^{1}$, Jose Carlos del Toro Iniesta$^{2}$, Joachim Woch$^1$, Achim Gandorfer$^1$, Johann Hirzberger$^1$, Wolfgang Schmidt$^3$, Thierry Appourchaux$^4$, Alberto Alvarez-Herrero$^5$ \and the SO/PHI team}

\affiliation{$^1$Max-Planck-Institut f\"ur Sonnensystemforschung, Justus-von-Liebig-Weg 3,
D-37077, G\"ottingen, Germany \\ email: {\tt solanki@mps.mpg.de} \\[\affilskip]
$^2$Instituto de Astrof\'{\i}sica de Andaluc\'{\i}a (IAA-CSIC), Apdo. de Correos 3004, E-18080 Granada, Spain \\[\affilskip]
$^3$ Kiepenheuer-Institut f\"ur Sonnenphysik, Sch\"oneckstr. 6, D-79104 Freiburg, Germany \\[\affilskip]
$^4$ Institut d'Astrophysique Spatiale, Universit\'e Paris Sud, B\^atiment 121, F-91405, Orsay, France \\[\affilskip]
$^5$ Instituto Nacional de T\'ecnica Aeroespacial, Carretera de Ajalvir, km. 4, E-28850 Torrej—n de Ardoz, Spain}

\pubyear{2008}
\volume{xxx}  
\pagerange{119--126}
\jname{Title of your IAU Symposium}
\editors{A.C. Editor, B.D. Editor \& C.E. Editor, eds.}
\begin{document}

\maketitle

\begin{abstract}
The {\em Solar Orbiter} is the next solar physics mission of the European Space
Agency, ESA, in collaboration with NASA, with a launch planned in 2018. The
spacecraft is designed to approach the Sun to within 0.28\,AU at perihelion of
a highly eccentric orbit. The proximity with the Sun will also allow its
observation at uniformly high resolution at EUV and visible wavelengths. Such
observations are central for learning more about the magnetic coupling of the
solar atmosphere. At a later phase in the mission the spacecraft will leave the
ecliptic and study the enigmatic poles of the Sun from a heliographic latitude
of up to 33$^\circ$.

A central instrument of {\em Solar Orbiter} is the Polarimetric and Helioseismic
Imager, SO/PHI. It will do full Stokes imaging in the Land\'e $g=2.5$ Fe\,I\,617.3\,nm
line. It is composed of two telescopes, a full-disk telescope and a
high-resolution telescope, that will allow observations at a resolution as
high as 200\,km on the solar surface. SO/PHI will also be the first solar
polarimeter to leave the Sun-Earth line, opening up new possibilities, such as
stereoscopic polarimetry (besides stereoscopic imaging of the photosphere and
stereoscopic helioseismology). Finally, SO/PHI will have a unique view of the
solar poles, allowing not just more precise and exact measurements of the
polar field than possible so far, but also enabling us to follow the dynamics
of individual magnetic features at high latitudes and to determine solar
surface and sub-surface flows right up to the poles.

In this paper an introduction to the science goals and the capabilities of
SO/PHI will be given, as well as a brief overview of the instrument and of
the current status of its development.

\keywords{Sun: magnetic fields, Instrumentation: high angular resolution,
Instrumentation: polarimeters, space vehicles: instruments, telescopes,
Sun: helioseismology, Sun: photosphere}

\end{abstract}

\firstsection 

\section{Introduction}
\label{se:intro}

{\em Solar Orbiter} is the first medium-class (M1) mission of the European
Space Agency's (ESA's) Cosmic Vision program and is expected to fly in 2018.
In addition to ESA, significant contributions will be provided by NASA. Within
the overall theme of ``How does the Sun create and control the heliosphere?"
the {\em Solar Orbiter} will address the following four top-level science questions:
\begin{itemize}
\item How and where do the solar wind plasma and magnetic field originate in the corona?
\item How do solar transients drive heliospheric variability?
\item How do solar eruptions produce energetic particle radiation that fills the heliosphere?
\item How does the solar dynamo work and drive connections between the Sun and the heliosphere?
\end{itemize}

The Polarimetric and Helioseismic Imager aboard {\em Solar Orbiter} (SO/PHI) will provide the magnetograms and helioseismic data needed by the mission to reach three of its four top-level science goals. It is a very challenging instrument to build. No space magnetograph has ever flown in such a difficult environment, with the spacecraft following a strongly elliptical trajectory, leading to huge thermal changes in the course of an orbit.

Here we present a first description of the science goals of SO/PHI as well as providing a brief description of the instrument and of its current status.

\section{Science goals}
\label{se:goals}

Among the four top-level science questions to be addressed by the
{\em Solar Orbiter} mission, SO/PHI will be crucial to answering three of
them, namely the first, the second, and the fourth mentioned in the
Introduction. These global questions flow down to three main scientific
requirements on {\em Solar Orbiter}:
\begin{enumerate}
\item investigate the links between the solar surface, corona, and inner heliosphere,
\item explore, at all latitudes, the energetics, dynamics and fine-scale structure of the Sun's magnetized atmosphere
\item probe the solar dynamo by observing the Sun's high-latitude field, flows and seismic waves.
\end{enumerate}

The coupling among the different parts of the solar atmosphere and the inner
heliosphere roots in the magnetic field. Any dynamic analysis of the solar
structures ---either magnetic or not--- needs to know the plasma velocities.
So does a study of flows and seismic waves. Therefore, to fulfill those
scientific requirements, SO/PHI is designed to measure the vector magnetic
field and the line-of-sight (LOS) velocity in the solar photosphere. The
extrapolation of the measured magnetic field to the corona and the inner
heliosphere will help to fulfill requirement
({\em a}). This task will be performed partly with high-resolution data,
partly with low resolution data (see next Section). Requirement~({\em b}) will
be undertaken with the help of both magnetic and velocity measurements,
mainly by using high-resolution data. While the investigations related to
these two first requirements will profit from the complementary information
of the other remote-sensing and in-situ instruments aboard the spacecraft,
requirement ({\em c}) will benefit almost exclusively from the
time series of LOS velocity and magnetic field measurements provided by SO/PHI,
so that this instrument will be paramount for addressing the fourth top-level science question of
{\em Solar Orbiter}.

A number of specific current scientific problems are expected to be addressed
in a unique way by SO/PHI. Some with SO/PHI working alone; many of them after
the combined action of SO/PHI with the other instruments. Thus SO/PHI will provide
the first maps of the polar vector magnetic fields from a vantage point well
above the ecliptic. It will also profit from a reduced solar apparent rotation
when following solar features, which will help in mitigating foreshortening
effects that especially harm the plasma velocity measurements. Joint
measurements with other {\em Solar Orbiter} instruments will address the
questions related to linkage science. SO/PHI will also be able to provide the
magnetic context to NASA's {\em Solar Probe Plus}. Due to its unique vantage
point, SO/PHI will also be capable of carrying out stereoscopic
imaging of brightness, magnetic fields
and velocities in combination with ground-based or near-Earth, space-borne
instruments. Getting rid of the 180$^{\circ}$ ambiguity inherent to the Zeeman
effect will then be achievable. An improved view of the vector velocity field
can also be expected from stereoscopy. Last, but not least, when in opposition
to the Earth, a complete 4$\pi$ view of the Sun's magnetic field will be
obtained, which will allow testing techniques such as far-side imaging.

Solar high-latitude science is still in its infancy
because optical measurements can only be carried out from the ecliptic.
Seen from there the solar pole never deviates by more than $7^{\circ}$ from the
plane of the sky, with the consequent strong foreshortening of features close
to the pole. Tsuneta et al. (2008) obtained the best
view so far of the polar magnetic fields with the {\em Hinode} satellite.
SO/PHI is expected to do considerably better and will help unravel the solar
rotation and the meridional flow at high latitudes. Accurate measurements of
the distribution and evolution of polar magnetic fields are important to test
current ideas about magnetic polarity reversals. Specifically, we need to know
how much magnetic flux is transported to high latitudes in the course of a
solar cycle (Wilson \etal\ 1990, Hoeksema \etal\ 2006). Studies of the origin
of the fast solar wind in polar coronal holes (e.g., Tu \etal\ 2005) are expected
to benefit from the vector magnetometry to be performed by SO/PHI from up to
33$^\circ$ above the ecliptic. The coverage of the whole range of solar
latitudes will allow us to investigate whether or not phenomena such as
convection (granulation and supergranulation) and quiet-Sun magnetoconvection (e.g.,
Requerey \etal, 2014,
and Kaithakkal \etal, 2015)), are the same at the
poles as near the equator.

The existence of a solar polar vortex was proposed theoretically by Gilman
(1976) and the first negative observational results were reported by Beckers
(1978). However, in recent years this phenomenon has become a matter of debate
again. Thus, Ye \& Livingston (1998) found indications
of a singularity that might be a vortex within 1$^{\circ}$ of the pole.
Benevolenskaya (2007) observed a sharp decrease of solar rotation at high
latitudes, although some other works seem to disagree and show a slow
decrease of the solar rotation rate close to the poles (e.g.,
Liu \& Zhao 2009 and references therein). SO/PHI will provide clear views of
the poles and thus will help to settle this debate.

Corbard \& Thompson (2002) reported evidence for a strong
gradient of the solar differential rotation in a layer close to the surface
and of its possible reversal at high latitudes. The possible effects of this
near-surface radial shear on flux-transport model dynamos is discussed by
Dikpati et al. (2002). SO/PHI measurements promise to be of considerable
relevance for determining the true level of shear at high latitudes.

New insights into the migrational flow pattern known as torsional oscillations, first
discovered by Howard \& LaBonte (1980), have been obtained whenever new instruments,
such as those of the Global Oscillations Network Group (Harvey et al. 1996), the
Michelson Doppler Imager (Scherrer et al. 1995), and the Helioseismic and Magnetic
Imager (Scherrer et al. 2012) have been used. Howe et al. (2013) have
combined results from all these three instruments to find changes in the
high-latitude branch of the torsional oscillations from one solar cycle to the
next, as modeled by Rempel (2012). A confirmation of such variations by
SO/PHI would definitively settle their existence, which has consequences for
the rotation pattern at high latitudes. Verification of the results of
Hathaway \& Upton (2014) on the variation of the meridional circulation over a
cycle and from one cycle to another is also needed, since they may have a
bearing on the strength of the following solar cycle.

\section{The SO/PHI instrument}

SO/PHI makes use of the Doppler and Zeeman effects in a single spectral line of
neutral iron at 617.3\,nm. The physical information is decoded from two-dimensional
filtergrams at six wavelength points within this line, while four polarization
states at each wavelength point are measured.

In order to obtain the abovementioned observables, SO/PHI is a diffraction
limited, wavelength tunable, quasi-monochromatic, polarization sensitive imager
with two telescopes, which (alternatively) feed a common filtergraph and focal plane array:

The High Resolution Telescope (HRT) provides a restricted FOV of 16.8\,arcmin squared and achieves a spatial resolution that, near the closest perihelion pass, will correspond to about 200\,km on the Sun. It is designed as a decentered Ritchey-Chr\'{e}tien telescope with a pupil of 140\,mm diameter. The two free-form aspheric mirrors are lightweight and made from ZERODUR, mounted in mirror cells from hardened INVAR and titanium.

The all-refractive Full Disk Telescope (FDT), with a FOV of 2$^\circ$ in diameter and a pixel size corresponding to 725\,km (at 0.28\,AU), provides a complete view of the full solar disk during all orbital phases. The FDT has an entrance pupil of 17.5\,mm and an effective focal length of 579\,mm. The FDT aluminium tube is mounted fixed in position to the back panel of the SO/PHI Optical Unit (OU) structure, while the front end of the tube is attached to the front panel using flex blades to allow thermal expansion in the axial direction.

These two telescopes are used alternatively and their selection is made by a feed selection mechanism.

The optics unit structure is common for both telescopes and consists of a combination of AlBeMet (an aluminum-beryllium alloy) and low expansion carbon-fibre reinforced plastic.

Both telescopes are protected from the intense solar flux by special
heat-rejecting entrance windows, which are part of the heat-shield assembly of
the spacecraft. These multilayer filters have more than 80\,\% \ transmittance
in a narrow notch around the science wavelength, while effectively blocking
the remaining parts of the spectrum from 200\,nm to the far infrared by
reflection. Only a fraction of the total energy is absorbed in the window,
which acts as a passive thermal element by emitting part of the thermal
radiation to cold space; emission of infrared radiation into the instrument
cavity is minimized by a low emissivity coating on the backside of the window
(acting at the same time as an anti-reflection coating for the science
wavelength). Thus the heat load into the instruments can be substantially
reduced, while preserving the high photometric and polarimetric accuracy of SO/PHI.

The filtergraph (FG) uses two key technologies with heritage from the Imaging Magnetograph eXperiment (IMaX, Mart\'{i}nez Pillet et al. 2011) onboard the successful Sunrise balloon-borne stratospheric solar observatory (Solanki et al. 2010; Barthol et al. 2011): a LiNbO$_3$ solid state etalon in a telecentric configuration selects a passband of 100\,m\AA\ width. Applying a high voltage across the crystal allows changing the refractive index and the thickness of the material, and thus tuning the passband in wavelength across the spectral line. A 3\,\AA \ wide prefilter acts as an order sorter for the Fabry-Perot channel spectrum.

Inherited as well from IMaX, the polarimetric analysis is perfomed by a Polarization Modulation Package (PMP) in each of the telescopes. Each PMP consists of two nematic liquid crystal retarders, followed by a linear polarizer as an analyzer. The liquid crystal variable retarders have been successfully qualified for the use in {\em Solar Orbiter} (see Alvarez-Herrero et al. 2011).

Both, the FG and the PMPs, are thermally insulated from the Optics Unit and actively temperature stabilized to within 0.045 and 0.5\,K, respectively. The opto-mechanical arrangement is designed to operate in a wide temperature range ($-20$ to $+50^{\circ}$C).

An internal image stabilization system in the HRT channel acts on the active
secondary mirror, which greatly reduces residual pointing error by the spacecraft to levels compatible with high resolution polarimetry. The error signal for the piezo-driven mirror support is derived from a correlation tracker camera inside the HRT.

The focal plane assembly is built around a 2048 by 2048 pixel Active Pixel Sensor (APS), which is especially designed and manufactured for the instrument. It delivers 10 frames per second, which are read out in synchronism with the switching of the polarization modulators.

\begin{figure}[h]
\hspace*{-12.9mm}\includegraphics[width=1.24\textwidth,clip=]{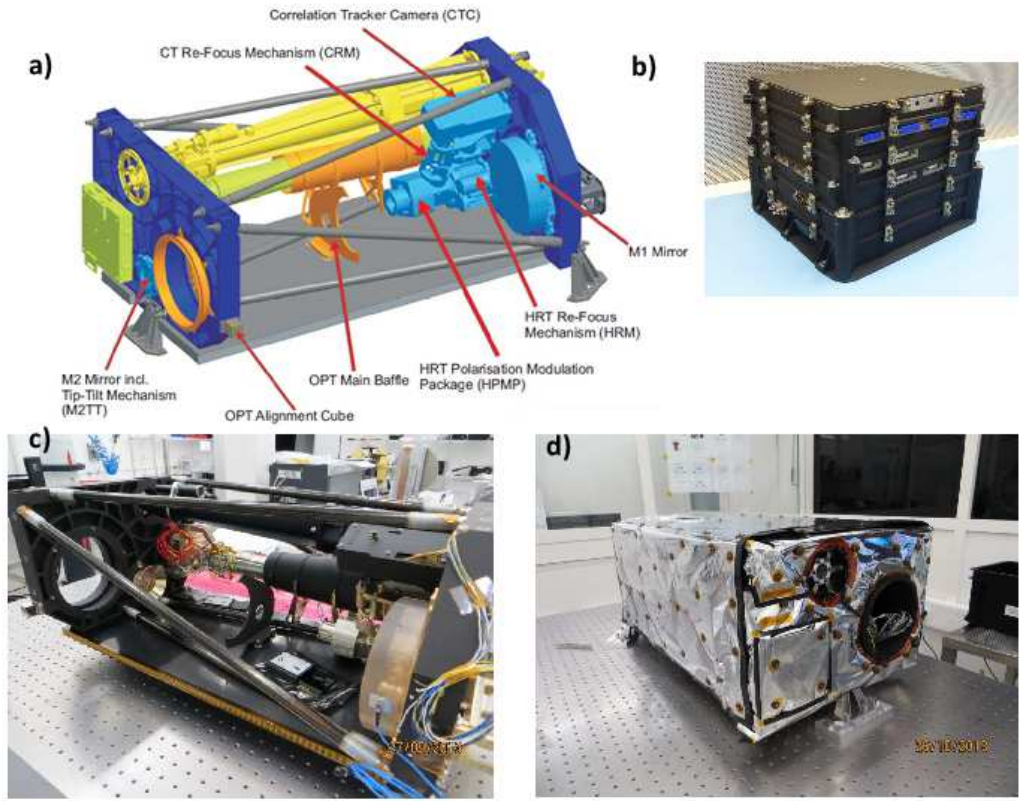}
\caption{Images of the SO/PHI instrument:
a) CAD representation of the SO/PHI Optics Unit with FDT tube (yellow) and HRT subsystems;
b) structural thermal model (STM) of the electronics unit;
c) STM of the optics unit without multilayer insulation (MLI);
d) same, but now with mounted MLI.}
\label{block}
\end{figure}

The extremely limited telemetry rate and the large amount of scientific
information retrieved from the SO/PHI instrument demand a sophisticated
on-board data reduction. The measurement technique of SO/PHI, i.e. the
determination of the full Stokes vector at several wavelengths, is well suited
to apply a robust and reliable procedure to obtain maps of
the three components of the vector magnetic field, the LOS velocity, and the
continuum intensity. To retrieve them, a non-linear, least-squares fitting
technique is used on board to numerically invert the radiative transfer
equation for polarized light. This non-linear and thus irreversible on-board
data analysis is facilitated by a powerful and adaptive Data Processing Unit
(see Fiethe \etal\ 2012). Although the variable thermal environment along the
orbit requires SO/PHI to retain full flexibility regarding the level of reduction
of its data products, the on-board data processing and the subsequent image
compression are necessary steps to achieving a reduction in data volume in
line with the limited telemetry.

\section{Status and outlook}

The SO/PHI instrument has passed its critical design review and is currently in the qualification phase. Flight model delivery is foreseen in 2016. After integration into the spacecraft and launch foreseen for 2018, a cruise phase of over three years will follow. At its end {\em Solar Orbiter} will be in its nominal science orbit with a perihelion at around 0.28\,AU close to the orbit of Mercury and in resonance with Venus. The nominal mission is set to last for four years, at the end of which, {\em Solar Orbiter} will have started to leave the ecliptic. It will move to still higher latitudes in the extened
phase of the mission, finally reaching 33$^\circ$. SO/PHI will carry out intensive observations of the Sun during three observing windows in each orbit and will provide lower cadence synoptic data between these windows.

\end{document}